\documentclass[prb,preprint,showpacs]{revtex4}

\usepackage{graphicx, epsfig}  
\usepackage{dcolumn} 
\def\be{\begin{equation}}
\def\ee{\end{equation}}
\def\bea{\begin{eqnarray}}
\def\eea{\end{eqnarray}}

\begin{document}

\title{Silver transport in Ge$_x$Se$_{1-x}$:Ag materials: \textit{ab initio} simulation of a solid electrolyte}

\author{De Nyago Tafen}
\email{tafende@helios.phy.ohiou.edu}

\author{D.A. Drabold}
\email{drabold@ohio.edu}

\affiliation{Department of Physics and Astronomy, Ohio University, 
Athens OH 45701, USA}

\author{M. Mitkova}
\affiliation{Center for Solid State Electronics Research, Arizona State University, 
Tempe, AZ 85287-6206, USA}

\pacs{61.43.Fs, 66.30.Dn, 66.30.Hs, 71.23.An, 71.23.Cq, 71.15.Mb }

\date{\today}

\begin{abstract}
In this paper, we present models of Ge-Se glasses heavily doped with Ag obtained from {\it ab initio} simulation and study the dynamics of the network with an emphasis on the motion of Ag$^+$ ions. The models are analyzed with partial pair correlation functions,  static structure factors
and novel wavelet techniques. The electronic properties are characterized by the electronic density of states and analysis of specific electronic eigenstates. As Ag content increases, the 
optical band gap increases.  Ag diffusion is observed directly from thermal simulation. The most diffusive Ag$^+$ ions move preferentially through low density regions of the network and the existence of well-defined trapping centers is confirmed.
Preliminary information about temperature dependence of trapping and release is provided.
\end{abstract}

\maketitle

\section{Introduction}
The chalcogenide glasses are preferred semiconducting materials for
applications. They have well-defined niches in fiber optics\cite{Aggarwal}, optical recording\cite{Mitkova}, phase change memory\cite{Ovshinsky}, and other technologies.  Ge-Se glasses have been particularly studied because 
of their ready glass formation, easy synthesis requirements,  and good chemical stability. The basic structural units are 
Se chains and Ge-Se tetrahedra which may combine in a variety of ways. Defects (including homopolar bonds) exist in
these systems as seen in both experiment and theory. 

The GeSe system was the first in which formation 
of an intermediate phase was demonstrated experimentally by Boolchand {\it et al}\cite{Boolchand} and further developed
theoretically by Thorpe, Jacobs, Chubnysky and Phillips\cite{Thorpe}. In binary Ge$_x$Se$_{1-x}$ 
glasses, the self-organized phase exists in the $0.20 < x < 0.254$ range, 
with glasses at $x < 0.20$ regarded as floppy while those with $x > 0.26$
stressed rigid.  Dynamic calorimetry measurements on the intermediate phase 
have led to the conclusion that such materials do not age\cite{Boolchand 2002}, a feature that may be of 
importance in application of these materials. 

Silver added to  chalcogenide glass hosts has attracted widespread interest in soft condensed matter 
science\cite{Mitkova 1999, Wang}.  The interest emerges in part from the extensive bulk glass 
forming tendency in the Ge-Se-Ag ternary, the spectacular enhancement (eight 
orders of magnitude) in electrical conductivity of glasses with Ag 
relative to the glassy chalcogenide hosts, and  from light-induced effects such as  photo-doping, 
photo-diffusion and  photo-deposition. Although the mobile ions in 
amorphous materials have been studied\cite{Angell} their 
detailed dynamics in amorphous hosts still constitutes one of the 
unsolved problems of solid state ionics.  The structure of Ge-Se-Ag glass has 
been investigated using several experimental 
methods, including X-ray diffraction\cite{Fischer, piarristeguy} neutron diffraction with 
isotopic substitution\cite{Lee}, EXAFS\cite{Oldale}, differential anomalous X-ray 
scattering (DAS)\cite{Dejus, Dejus 1992, Westwood} and Modulated Differential Scanning Calorimetry 
(MDSC) and Raman spectroscopy\cite{Mitkova 1999, Wang}. Despite this impressive database, the 
structure of the ternary Ge-Se-Ag glasses has not yet been completely 
determined. There continues to be a debate on basic aspects of the glass 
structure (i.e. homogeneity and Ag coordination) especially for Se rich 
glasses with more than 67\% Se.  

Experimental evidence for macroscopic phase separation in these materials has come from MDSC results, which indicate bimodal glass transition temperatures\cite{Wang} . In these experiments, one T$_g$  is independent of glass composition, and identified with a Ag$_2$Se glass phase\cite{pbnature}, while the second T$_g$ that varies with glass composition is related to the  Ge-Se backbone. For the time scales (and possibly also lengths scales) of our simulations, these effects do not emerge, but are an interesting challenge for the future.

In a recent Letter\cite{tafdad05}, we briefly reported 
the motion of Ag ions in glassy chalcogenide hosts, and demonstrated the
existence of ion trapping centers,  which are important for relaxation processes 
in disordered systems\cite{jcp}. Here,  we provide detailed information 
about structural and electronic properties, and also 
give new information about temperature dependence
of the trapping, and also the geometry of the traps. In this paper, we have focused upon Ag-doped glasses containing Ge 25
at.\% and Se 75 at.\%  (we later call this GeSe$_3$). This composition is near the intermediate phase 
(slightly into the stressed-rigid phase). To our knowledge, this is the first {\it ab initio} simulation of these materials.

The rest of this paper is organized as follows. In Section II, we describe the simulation procedure to fabricate the atomistic models, discuss the {\it ab initio} total energy functional and force code used and other approximations. In Section III, we describe the structural properties using conventional measures such as static structure factors,  and also apply a novel wavelet method to explore intermediate range order. Electronic properties are briefly discussed in Section IV, and Section V is concerned with the dynamics of the Ag$^+$ ions in the amorphous matrix. 

\section{Model Generation}

\subsection{Energy Functional and Interatomic Forces}
For the simulations reported in this paper, we use FIREBALL2000 developed by Lewis and coworkers\cite{Lewis}. This code is an approximate {\it ab initio} density functional approach to the electronic structure, total energies and forces based upon pseudopotentials and a real-space local basis of slighty-excited pseudoatomic orbitals to represent the Kohn-Sham functions. The method uses separable pseudopotentials, and allows the use of double-zeta numerical basis sets and polarization orbitals. The calculation is undertaken entirely in real space, which provides substantial computational efficiency. The 
exchange-correlation energy was treated within the LDA, using the results of Ceperley and Alder\cite{ceperley}, as 
interpolated by Perdew and Zunger\cite{perdew} (more intricate gradient corrected functionals are available if needed). The pseudopotential and pseudoatomic wave functions were generated in the Troullier-Martins form\cite{troullier} employing the 
scheme of Fuchs and Scheffer\cite{fuchs}.  Hamiltonian and overlap matrix elements are precalculated on a numerical grid and the specific values needed for a particular instantaneous conformation are extracted from the tabulated values via interpolation.
Naturally, the integral tables need to be generated only once, for a given set of atomic species, rather than performing quadratures ``on the fly'' during a MD run. 
\subsection{Model Formation}

The models described here were generated using the melt quenching method. We began by randomly placing atoms in a cubic supercell according to the desired composition [for (GeSe$_3$)$_{0.90}$Ag$_{0.10}$ 54 germanium atoms, 162 selenium atoms and 24 silver atoms; for (GeSe$_3$)$_{0.85}$Ag$_{0.15}$ 51 germanium atoms, 153 selenium atoms and 36 silver atoms] with the minimum acceptable distance between atoms 2 \AA. The size of the cubic cells was chosen to make the density of these glasses close to experimental data. The box size of the 240 atom supercell of (GeSe$_3$)$_{0.90}$Ag$_{0.10}$ and (GeSe$_3$)$_{0.85}$Ag$_{0.15}$ are respectively 18.601 \AA and 18.656 \AA with corresponding density\cite{piarristeguy} 4.98 g/cm$^3$ and 5.03 g/cm$^3$. The structures were annealed and we obtained well thermalized melts at 4800K. We took three steps to cool the cells. First, the cells were equilibrated to 1100 K for 3 ps; then they were slowly cooled to 300 K over approximately 5 ps. The MD time step was 2.5fs. Simple velocity rescaling was used for the dissipative dynamics. In the final step, the cells were steepest descent quenched to 0K and maximum forces smaller in magnitude than 0.02 eV/\AA. All calculations were performed at constant volume using the $\Gamma$ point to sample the Brillouin zone to compute energies and forces.

The use of MD to quench a model liquid is natural and \textit{in principle} completely general, since it superficially mimics the process of glass formation. It has been utilized for structural calculations for GeSe$_2$\cite{Cappelletti}, As$_2$Se$_3$\cite{Jun}, Ge-Se-Ag\cite{Iyetomi} and other systems (though there are sometimes hints that there is too much ``liquid-like'' character frozen into the resulting models). With accurate force calculations, simulated melt quenching has produced  disappointing results for liquids such as GeSe\cite{Roon} and ternary glasses such as As-Ge-Se\cite{comment}. We believe that the success with (GeSe$_3$)$_{1-x}$Ag$_x$ ($x$=0.10, 0.15) glasses is connected to the fact we are in a weakly overconstrained glass forming part of the vibrational phase diagram\cite{mikejim}. The ``cook and quench'' method is less  effective in the highly overconstrained regime.  It is likely that hybrid schemes mixing experimental information and {\it ab intio} simulation (such as the ``ECMR'' method\cite{ecmr}) would be most effective in this composition regime.

\section{Structural Properties}

\begin{table*}
\caption{\label{tab:table1} Basic short range order parameters in models {\it g}-(GeSe$_3$)$_{0.90}$Ag$_{0.10}$ ($x$=0.10) and {\it g}-(GeSe$_3$)$_{0.85}$Ag$_{0.15}$ ($x$=0.15) and compared to available data. $r_1$ is the average bond length, $\bar{n}$ the average coordination number, and $r_2$ the second nearest neighbor distance.}
\begin{ruledtabular}
\begin{tabular}{lccccccc}
&\multicolumn{3}{c}{This Work}&\multicolumn{3}{c}{Experiment (Ref. \cite{piarristeguy})}\\
Glass & $r_1$ ($\pm$0.01 \AA) & $r_2$ ($\pm$0.01 \AA) & $\bar{n}$ ($\pm$0.01) & $r_1$ ($\pm$0.02 \AA) & $r_2$ ($\pm$0.05 \AA) & $\bar{n}$\\
\hline
$x$=0.10 & 2.47 & 3.78 & 2.525 & 2.46 & 3.80 & 2.471$\pm$0.002\footnotemark[1]  2.543$\pm$0.100\footnotemark[2] \\
$x$=0.15 & 2.48 & 3.82 & 2.775 & 2.48 & 3.82 & 2.855$\pm$0.002\footnotemark[1]  2.852$\pm$0.100\footnotemark[2] \\
\end{tabular}
\end{ruledtabular}
\footnotetext[1]{Analytical calculation}
\footnotetext[2]{Monte Carlo method}
\end{table*} 

\subsection{Short range order and defects}
We define short range order (SRO) as a length scale not longer than the second nearest neighbor distance. Table \ref{tab:table1} gives an overview of the SRO in \textit{g}-(GeSe$_3$)$_{0.90}$Ag$_{0.10}$ and \textit{g}-(GeSe$_3$)$_{0.85}$Ag$_{0.15}$. The average bond length and average coordination number closely agree with the available data\cite{piarristeguy}. With increasing Ag concentration, the average coordination number increases to 2.8 in \textit{g}-(GeSe$_3$)$_{0.85}$Ag$_{0.15}$ from 2.51 in \textit{g}-GeSe$_3$\cite{Petri}. Also listed is the second nearest neighbor distance. Table \ref{tab:table2} lists the average bonding distances of different possible bonds present in the model.  Different values of the Ag-Ag bond distance have been proposed. Using the differential anomalous scattering (DAS), Westwood \textit{et al}\cite{Westwood} obtained a value of 3.35 $\AA$ for the Ag-Ag distance, a bit longer than our observation or other data\cite{piarristeguy}.
Both models contain structural defects. In addition to the normally coordinated Ge$_4$ and Se$_2$: Ge$_3$, Se$_3$ and Se$_1$ are present in both models. Table \ref{tab:table3} summarizes the statistical distribution of the main structural components. A look at the table shows that Ge-Se, Se-Se, Ag-Se and Ag-Ge correlations depend upon $x$ with additional Ag modifying the Ge-Se and Se-Se bonding.  By integrating the partial pair correlation function g$_{\alpha\beta}$(r) (Fig. \ref{fig4}), we estimate the average coordination number of Ag in the models. We found an average coordination of 2.0 and 2.9 in \textit{g}-(GeSe$_3$)$_{0.90}$Ag$_{0.10}$ and \textit{g}-(GeSe$_3$)$_{0.85}$Ag$_{0.15}$ respectively. We note that this is hard to estimate without ambiguity because of the lack of a well-defined deep minimum in $g(r)$ after the first peak especially for the $x=0.15$ model.

When discussing the coordination of Ag one can follow the arguments of 
Kastner\cite{Kastner} who was the first to describe the bonding of metal atoms in 
chalcogenide glasses. However, we have to consider the fact that in 
addition to the covalent bond that is expected to form with the 
chalcogens, Ag offers three empty $s-p$ orbitals and is surrounded by the 
lone-pair electrons of its chalcogen neighbors. The latter offer the 
opportunity for the formation of up to three coordinate bonds. A 
coordinate bond is similar to a covalent bond and has similar strength 
but the bonding electrons are supplied by one bonding partner (the 
chalcogen atom)\cite{Cotton}. As a result, the lowest energy-bonding configuration for Ag in 
chalcogenide glasses is an overall neutral complex with Ag positively 
charged with the negative charge located on neighboring chalcogen or 
chalcogens\cite{Fritzsche}. The coordination of Ag thereby can vary,  but the 
expected value would be in average close to three - with one covalent 
bond and up to 2 coordinated bonds. The opportunity for four-fold 
coordination also exists as there is one more free $s-p$ orbital at the Ag 
atom but evidently this does not satisfy the requirements of the lowest 
energy configuration and the electronegativity of the entire complex in 
the presence of other cations in the system so that the probability for 
this type of bonding is lower. Our simulations appear to be consistent with these
chemical considerations, as all of the  Ag is two-fold in the 10\% model, with 
the addition of three-fold Ag in the 15\% Ag model.

\subsection{Intermediate range order}
\begin{figure}
\includegraphics[width=3.0 in, height=3.0 in, angle=0]{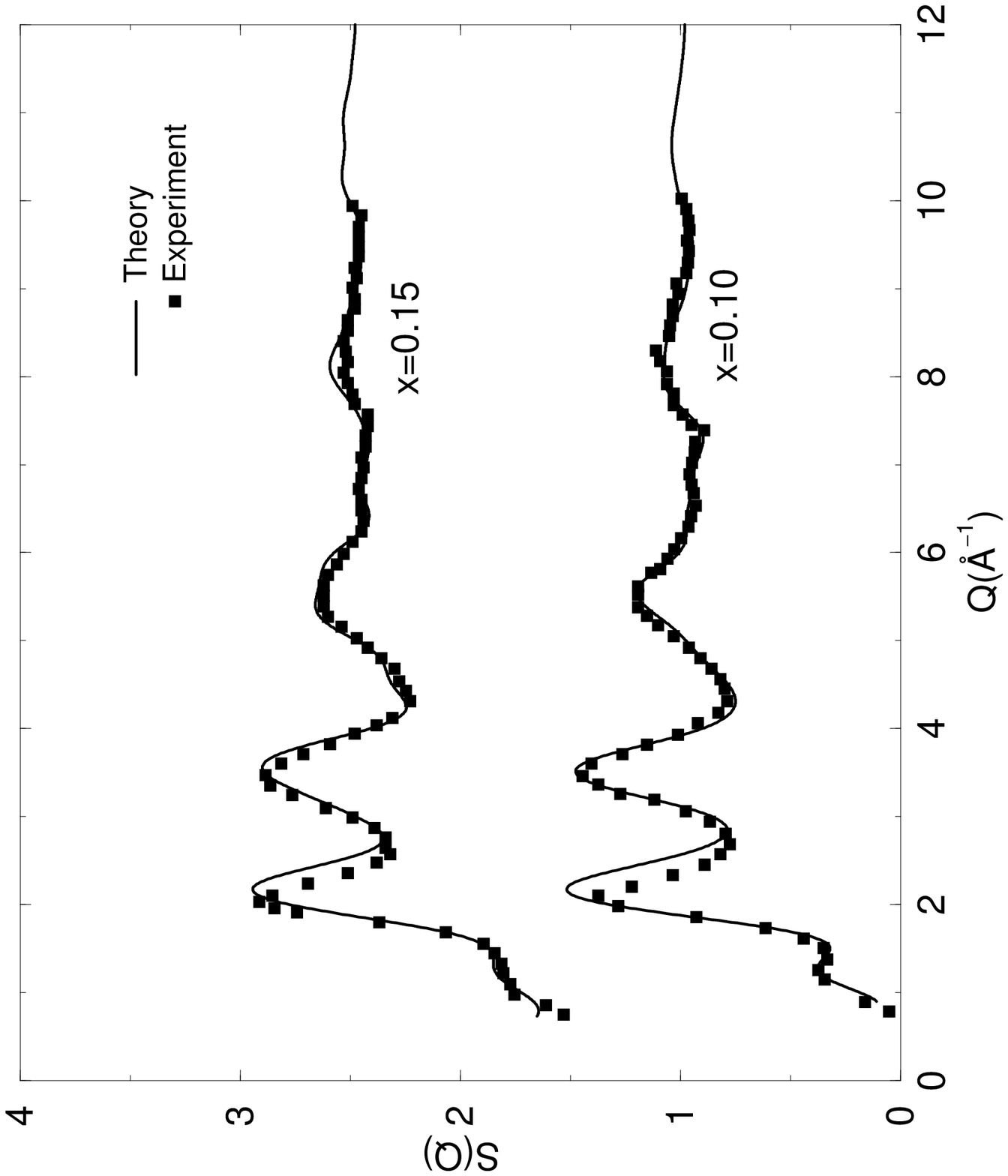}
\caption{Calculated total structure factor S(Q) of (GeSe$_3$)$_{0.90}$Ag$_{0.10}$ and (GeSe$_3$)$_{0.85}$Ag$_{0.15}$ glasses
compared to experiment\cite{piarristeguy}.
 \label{fig1}}
\end{figure}

\begin{figure}
\includegraphics[width=3.0 in, height=3.0 in, angle=0]{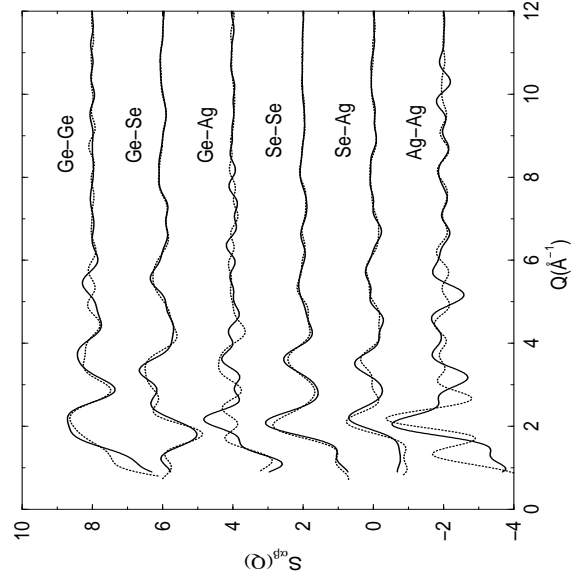}
\caption{Partial structure factors S$_{\alpha\beta}$(Q) of (GeSe$_3$)$_{0.90}$Ag$_{0.10}$ (solid lines) and (GeSe$_3$)$_{0.85}$Ag$_{0.15}$ (dotted lines) glasses.
\label{fig2}}
\end{figure}

Fig. \ref{fig1} shows the calculated static structure factors for (GeSe$_3$)$_{0.90}$Ag$_{0.10}$ and (GeSe$_3$)$_{0.85}$Ag$_{0.15}$ and the comparison with the experimental data from ref. \cite{piarristeguy}. As we did not include \textit{a priori} information in the model formation process, the fact that the peak positions and spectral weight of S(Q) agree well with experimental data is encouraging. The third and fourth peaks are a result of the short range order in the models.  The position of the second peak does not depend strongly on the Ag concentration (though its width does). This peak is located at Q$\cong$2.09 \AA$^{-1}$ in both models. By contrast the third peak intensity decreases with increasing Ag. This peak is located at Q$\cong$3.49 \AA$^{-1}$ in (GeSe$_3$)$_{0.90}$Ag$_{0.10}$ and at Q$\cong$3.41 \AA$^{-1}$ in (GeSe$_3$)$_{0.85}$Ag$_{0.15}$; whereas in GeSe$_3$ (\textit{x}=0, concentration of Ag in (GeSe$_3$)$_{1-x}$Ag$_x$) this peak appears near\cite{piarristeguy, Petri} 3.53 \AA$^{-1}$. In Fig. \ref{fig2} the partial structure factors show that S$_{GeGe}$(Q) and S$_{GeSe}$(Q) tend to cancel each other. 

\begin{figure}
\includegraphics[width=4.0 in, height=4.0 in, angle=0]{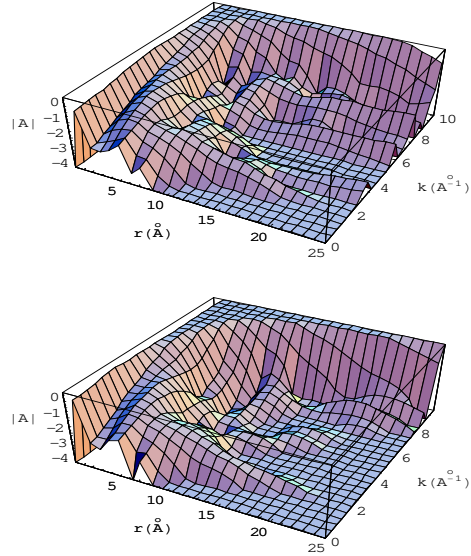}
\caption{(Color Online) Three-dimensional $r-k$ diagrams of the instantaneous amplitude A of the continuous wavelet transform of the experimental structure factors\cite{piarristeguy} for (GeSe$_3$)$_{0.90}$Ag$_{0.10}$ (top) and (GeSe$_3$)$_{0.85}$Ag$_{0.15}$ (bottom) glasses, using the methods of Harrop and coworkers\cite{harrop}. \label{fig3}}
\end{figure}

\begin{figure}
\includegraphics[width=3.0 in, height=3.0 in, angle=0]{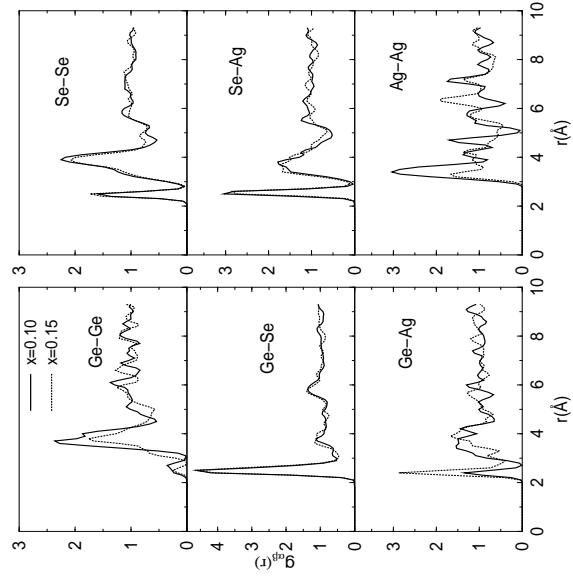}
\caption{Partial pair correlation functions g$_{\alpha\beta}$(r) of (GeSe$_3$)$_{0.90}$Ag$_{0.10}$ (solid lines) and (GeSe$_3$)$_{0.85}$Ag$_{0.15}$ (dotted lines) glasses. 
\label{fig4}}
\end{figure}

Both models reveal a feature in S(Q) near 1.07 \AA$^{-1}$. This is a harbinger of a First Sharp Diffraction Peak (FSDP), that becomes explicit in more Ge-rich materials.  Piarristeguy \textit{et al}\cite{piarristeguy} show that this peak varies as a function of Ag content. As Ag concentration increases, the FSDP decreases due to a change of the IRO.  Moreover, Ag disturbs the GeSe$_{4/2}$ network, and leads to the fragmentation of GeSe$_{4/2}$ tetrahedrons. From the partial structure factors it is apparent that the ``proto-FSDP" has contributions from all of the partials. To elucidate the intermediate range order more quantitatively, we have used the wavelet-based methods of Harrop and coworkers\cite{Harrop}. This is a promising scheme for interpreting structure factor data based upon continuous wavelet transforms\cite{waveletnote} (CWT). In Fig. \ref{fig3}, we illustrate the results of the wavelet analysis extracted from the experimental data\cite{piarristeguy}. The most obvious feature is the extended range real-space correlations associated with the diffraction peak near 3.5\AA$^{-1}$ for the 10\% Ag model. This is connected to the narrowness of the peak for the 10\% glass relative to the 15\% material: in the latter case the correlations disappear by about 15\AA$^{-1}$, whereas the correlations extend to at least 25\AA for the 10\% glass. A similar state of affairs also accrues for the peak near 5.5\AA$^{-1}$ and for a similar reason. This work emphasizes that simple associations of the reciprocal of the peak position of the FSDP to real-space length scales is misleading, particularly if the peak is narrow as discussed by Uchino and coworkers in studies of silica glass\cite{Harrop}.

In Fig. \ref{fig4} we plot the partial pair correlation function of both models. The Ge-Se and Se-Se pairs provide the dominant contribution to the first shell of the pair correlation function g(r) whereas Ag-Se contribute to the second peak, Ag-Ag to the third and Se-Se (second nearest neighbors) to the fourth peak.
     
\begin{table}
\caption{\label{tab:table2} Nearest-neighbor distances in (GeSe$_3$)$_{0.90}$Ag$_{0.10}$ ($x$=0.10) and (GeSe$_3$)$_{0.85}$Ag$_{0.15}$ ($x$=0.15) glasses.}
\begin{ruledtabular}
\begin{tabular}{lccccccc}
  & Correlation & Distance ($\pm$0.03 \AA) & Ref.\onlinecite{piarristeguy} ($\pm$0.05 \AA) \\ \hline
$x$=0.10 & Ge-Ge & 2.38 &   \\
   & Ge-Se & 2.37 & 2.37 \\
   & Ge-Ag & 2.35 &      \\
   & Se-Se & 2.36 & 2.37 \\
   & Ag-Se & 2.63 & 2.67 \\
   & Ag-Ag & 3.10 & 3.05 \\
\\
$x$=0.15 & Ge-Ge & 2.38 &   \\
   & Ge-Se & 2.37 & 2.37 \\
   & Ge-Ag & 2.35 &      \\
   & Se-Se & 2.36 & 2.37 \\
   & Ag-Se & 2.63 & 2.67 \\
   & Ag-Ag & 3.21 & 3.05 \\
\end{tabular}
\end{ruledtabular}
\end{table} 

\begin{table*}
\caption{\label{tab:table3} The statistical distribution of the main structural components in \textit{g}-(GeSe$_3$)$_{0.90}$Ag$_{0.10}$ ($x$=0.10) and \textit{g}-(GeSe$_3$)$_{0.85}$Ag$_{0.15}$ ($x$=0.15) models. Percentage of a given component in the total configuration is given in parentheses. Here the subscript number indicates the coordination number. For example, Ag$_2$ means the two-bonded Ag sites.}
\begin{ruledtabular}
\begin{tabular}{lccccccc}
  & Ge$_4$ & Ge$_3$ & Se$_2$ & Se$_3$ & Se$_1$ & Ag$_2$ & Ag$_3$ \\ \hline
$x$=0.10 & 35 (65\%) & 17 (31.5\%) & 86 (53.1\%) & 56 (33.9\%) & 17 (11.8\%) & 24 (100\%) & - \\
$x$=0.15 & 34 (66.7\%) & 13 (25.5\%) & 85 (55.6\%) & 57 (37.2\%) & 11 (7.2\%) & 31 (86.1\%) & 4 (11.1\%) \\
\\
 & Ge-Se & Se-Se & Ag-Se & Ag-Ge \\ \hline
$x$=0.10 & 64\% & 21\% & 13\% & 1.5\% \\
$x$=0.15 & 57.1\%& 19.2\%& 18.2\%& 5.2\%\\
\end{tabular}
\end{ruledtabular}
\end{table*} 

\section{Electronic Properties}
Having studied structural properties, we now briefly analyze the electronic properties of our models.  The electronic density of states (EDOS) of both models are calculated and analyzed by the inverse participation ratio (IPR), which we denote by $\cal{I}$. The EDOS are obtained by summing suitably broadened Gaussians centered at each eigenvalue. The IPR
\begin{displaymath}
 {\cal{I}} (E)=N \sum_{n=1}^N q(n,E)^{2} 
\end{displaymath}
determines the spatial localization of electronic eigenvalues. Here $N$ is the number of atoms in the model and $q(n,E)$ is the Mulliken charge localized on atomic site $n$ in a certain eigenstate $E$. Hence, $\cal{I}$ is a measure of the inverse number of sites involved in the state with energy E. For a uniformly extended state, the Mulliken charge contribution per site is uniform and $\cal{I}$(E)=1/N. For an ideally localized state, only one atomic site contributes all the charge and $\cal{I}$(E)=1. Therefore a larger value of $\cal{I}$ means that the eigenstate is more localized in real space.

\begin{figure}
\includegraphics[width=3.0 in, height=3.0 in, angle=0]{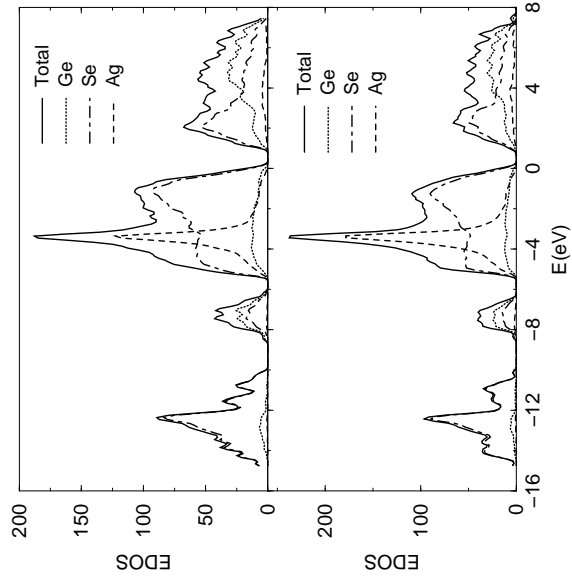}
\caption{Electronic density of states and species projected electronic density of states for Se, Ge, and Ag for (GeSe$_3$)$_{0.90}$Ag$_{0.10}$ (top panel) and (GeSe$_3$)$_{0.85}$Ag$_{0.15}$ (bottom panel) glasses. \label{fig5}}
\end{figure}

In Fig. \ref{fig5} we report the EDOS and the species-projected density of states of (GeSe$_3$)$_{0.90}$Ag$_{0.10}$ and (GeSe$_3$)$_{0.85}$Ag$_{0.15}$ glasses (our electronic eigenvalues have been shifted in order to place the valence band edge eigenvalue at zero). It should be noted that the spectra of both models are similar and closely related to the EDOS of Ge$_x$Se$_{1-x}$ ($x>0.15$). With the addition of Ag into \textit{g}-GeSe$_3$, an intense peak, due to the Ag 4$d$ electrons appears at about -3.47 eV as shown in Fig. \ref{fig5}. The valence band exhibits three features. The two lowest bands between -14.8 eV and -7.0 eV originate from the atomic 4$s$-like states of Ge and Se partially hybridized to form bonding states to Ag atoms. The next band lying between -7.0 and 0.0 eV contains $p$ like bonding states of Ge and Se and $d$ like bonding states of Ag. The peak in the topmost valence region is due to the lone-pair 4$p$ electrons of Se atoms. The $\Gamma$ point optical gaps of \textit{g}-(GeSe$_3$)$_{0.90}$Ag$_{0.10}$ and \textit{g}-(GeSe$_3$)$_{0.85}$Ag$_{0.15}$ are respectively of the order of 1.20 and 1.26 eV\cite{gap}.  As the Ag content increases, the optical band gap increases. To our knowledge, experimental information about the EDOS of both systems is unavailable, so the curve in Fig. \ref{fig5} is actually a \textit{prediction}. This is an interesting contrast to the work of Simdyankin and coworkers\cite{sim}, who show that for {\it low} concentrations of Cu,  the gap {\it decreases} with addition of Cu in AsS and AsSe glasses. Care is needed in comparing these results since the hosts and transition metals are different,  and our models have far higher metal content.

\begin{figure}
\includegraphics[width=3.0 in, height=3.0 in, angle=0]{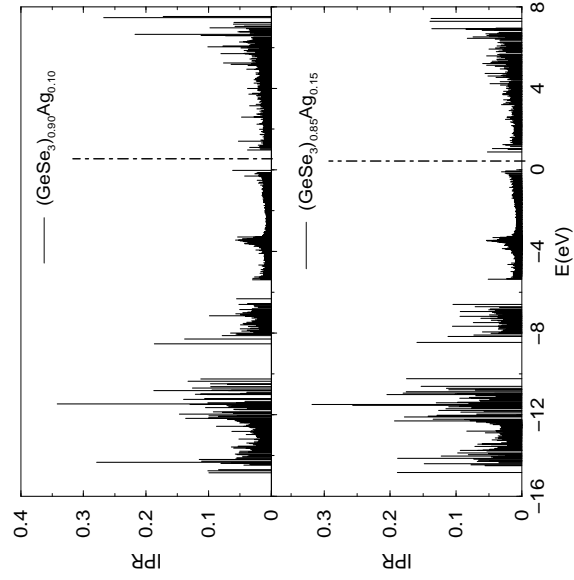}
\caption{Inverse participation ratio of (GeSe$_3$)$_{0.90}$Ag$_{0.10}$ (top panel) and (GeSe$_3$)$_{0.85}$Ag$_{0.15}$ (bottom panel) glasses. The vertical dot-dashed line indicates the position of the Fermi level.\label{fig6}}
\end{figure}

In order to connect localized eigenstates to particular topological/chemical regularities we plot in Fig. \ref{fig6} the IPR in the band gap region. We found that the localization in the valence band is mainly due to Se atoms and in the conduction band to Ge atoms in both models. A close look at the localized states at the band edges shows that the localized states at the top of the valence band of \textit{g}-(GeSe$_3$)$_{0.90}$Ag$_{0.10}$ are mostly associated with two- and three-fold coordinated Se atoms with homopolar bonds, whereas the localization at the conduction band edge arises from overcoordinated Se associated with homopolar bonds and four-fold coordinated Ge connected to Se atoms involved with Se-Se homopolar bonds. By contrast, the top of the valence band of \textit{g}-(GeSe$_3$)$_{0.85}$Ag$_{0.15}$ is quite extended; the conduction band edge shows very few localized states due to the overcoordinated Ge and Se. This explains the results of Kawasaki\cite{Kawasaki} showing dominance of the ionic conductivity related to Ag$^+$ ions at these particular compositions. 

\section{The Dynamics of Silver Ions}
\begin{figure}
\includegraphics[width=3.0 in, height=3.0 in, angle=0]{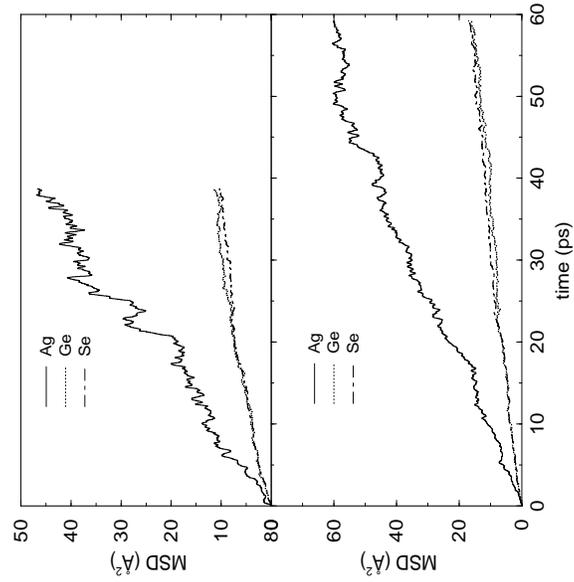}
\caption{Mean square displacement for all of the atomic species in (GeSe$_3$)$_{0.90}$Ag$_{0.10}$ (top panel) and (GeSe$_3$)$_{0.85}$Ag$_{0.15}$ (bottom panel) glasses simulated at T=1000 K.\label{fig7}}
\end{figure}

An outstanding feature of the materials is the high mobility of the silver in the complex host. 
The chemical explanation for the high diffusivity of 
Ag is its high quadrupolar deformability\cite{Fritzsche}. To ``mine"  information from our MD simulations, we begin
by computing the mean-square displacement (MSD) functions for all of the atomic constituents:
\be
\langle r^2(t)\rangle_\alpha= \frac{1}{N_\alpha}\sum_{i=1}^{N_\alpha} \langle {| \bf{r}_i (t)- \bf{r}_i} (0) |^2\rangle,
\ee
where $\langle \rangle$ is an average over MD simulation time and the sums are over particular atomic species, $\alpha$. The MSD were calculated for Ge, Se and Ag ions for both models and are shown in Fig. \ref{fig7}. The MSD of Ag ions increases rapidly with time, whereas that of the Ge and Se show a very slight slope.  We chose a temperature of 1000 K to illustrate the diffusion, and later discuss behavior at selected lower temperatures.

\begin{figure}
\includegraphics[width=3.0 in, height=3.0 in, angle=0]{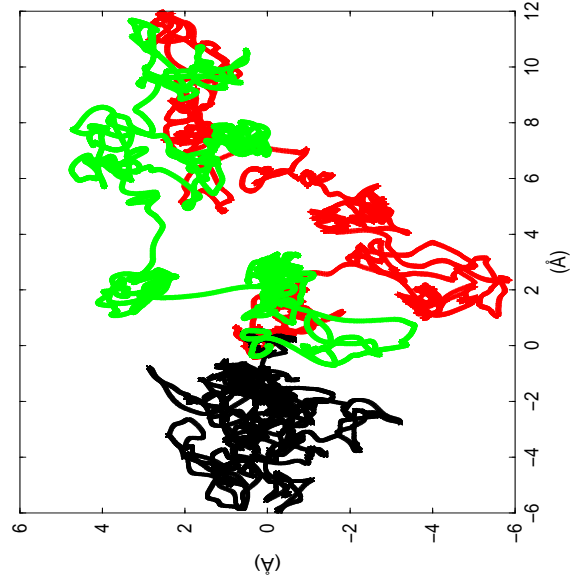}
\caption{(Color Online) Trajectories of the most (green and red) and least (black) mobile Ag atoms in (GeSe$_3$)$_{0.85}$Ag$_{0.15}$ glass (T=1000 K).  \label{fig8}}
\end{figure}

\begin{figure}
\includegraphics[width=3.0 in, height=3.0 in, angle=0]{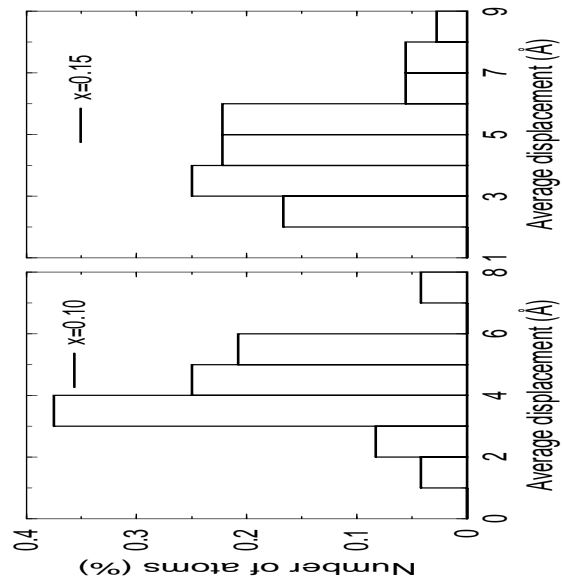}
\caption{Histogram of RMS displacements of Ag ions in both models. Left panel, 10\% Ag, right panel 15\% Ag. 39ps of constant temperature MD at 1000K was used to accumulate these statistics. \label {fig9}}
\end{figure}

To explain the mechanism of diffusion of silver, we examine the trajectories of these particles. We obtained 2.5$\times$10$^4$ steps of time development, for a total time of 62.5 ps and a fixed temperature of 1000 K. Fig. \ref{fig8} illustrates 2D projections of trajectories of the most and least mobile Ag atoms in the $x=0.15$ model.  We notice that for short times, the MSD of the most mobile atoms increases due to the diffusive motion of Ag. At intermediate times, the atoms may be trapped in a cage formed by their neighbors, and at the longest times we can explore, they can escape such traps and diffuse again. Thus, our trajectories can largely be separated into vibration around stable trapping sites and hops between such sites.  In both glasses, a fraction of Ag atoms move large distances (see Fig. \ref{fig9}). In (GeSe$_3$)$_{0.90}$Ag$_{0.10}$, about 91.67 \% of silver atoms move an average distance greater than 2.5 \AA for a time scale of 39 ps. Among them 25 \% have an average displacement greater than 5 \AA. By contrast only 4.2 \% of Ag atoms move less than 2 \AA. On the other hand, about 89 \% of Ag atoms move on an average distance greater than 2.5 \AA in (GeSe$_3$)$_{0.85}$Ag$_{0.15}$ for the same period of time. 36.1 \% of those atoms have an average displacement greater than 5 \AA. The most mobile Ag atoms move on an average distance of 8.2 \AA. These numbers illustrate the high ionic mobility of Ag ions in these complex glasses and are suggestive for a significant contribution of correlated hops of Ag$^{+}$ in the diffusion process.

Based on these trajectories, thermal transport coefficients such as diffusion coefficients can be evaluated. Typically one uses either the Green-Kubo formula \cite{Chandler} where the time-dependent velocity autocorrelation (VAC) is integrated or the Einstein relation \cite{Chandler} is employed, and the MSD is differentiated with respect to time. Since transport coefficients are equilibrium properties, the system must be properly thermalized before the transport properties can be estimated.  Our simulations show that at time t$>$4ps the systems are well equilibrated . The Einstein relation for self-diffusion reads:
\be
\langle|\bf{r}(t)-\bf{r}(0)|^2\rangle=6Dt+C
\ee
where C is a constant,  $D$ is the self-diffusion coefficient and $\langle|\bf{r}(t)-\bf{r}(0)|^2\rangle$ is the mean-square distance from initial position at time $t$, averaged over atoms of a given species. Direct simulation of the atomic trajectory and simple fitting yields C and D, and,  in particular,  estimates for the self-diffusion coefficient of Ag, $D_{Ag}$. The estimated values of $D_{Ag}$ as a function of temperature are listed in Table \ref{tab:table4}. These results are qualitatively reasonable when compared to the recent (room temperature) experiments of Ure$\tilde{n}$a \textit{et al}\cite{Urena} with the appropriate exponential activation factor included. The probability of correlated motion of Ag$^{+}$ will of course increase with increasing Ag concentration in agreement with decrease of the activation energy for conductivity\cite{Gutenev}.

\begin{figure}
\includegraphics[width=3.0 in, height=3.0 in, angle=0]{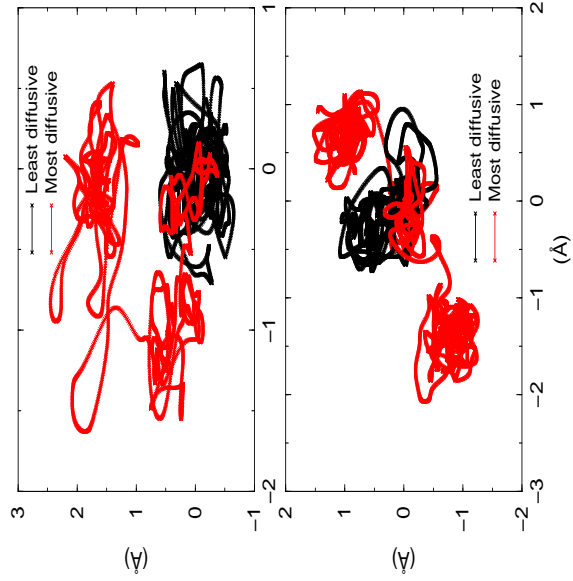}
\caption{(Color Online) Trajectories of the most and least mobile Ag atoms in (GeSe$_3$)$_{0.90}$Ag$_{0.10}$ (top panel) and (GeSe$_3$)$_{0.85}$Ag$_{0.15}$ (bottom panel) glasses (T=640 K). \label{fig10}}
\end{figure}

The dynamics is sensitive to the temperature. We performed additional simulations at temperatures ranging from 640K to 1000K. In Fig. \ref{fig10} we illustrate the hopping, which is qualitatively like the high temperature hopping, and the traps are very well defined. We note that even the most mobile Ag ions spend a substantial time in the traps, and appear to hop very efficiently (quickly) between traps.

To further study the trapping centers\cite{Phillips} we also obtained estimates of trap sizes and trap lifetimes as a function of temperature. In our calculations, we consider only particles that experience more than one traps. To calculate the size of the traps we enclose the particle trajectories of each trap in a sphere of radius $r_{tr}$ centered on the average position of the particle in the trap. Then, we determine the displacement of the particle trajectories in the trap with respect to the average trajectory, and  average over all the displacements. The trap size $r_{tr}$ is then obtained by averaging over different traps. Knowing the trap sizes, their lifetimes can be easily determined. In Table \ref{tab:table4}, we give an estimate of the trap sizes and trap lifetimes as a function of temperature. As T increases, the trap sizes increase and their lifetimes decrease. The averaged trap radii range from about 0.8 to more than 1.8 \AA, and the averaged lifetimes extend from about 2.5 to more than 5 ps. The individual trap radii can be as large as 2.4 \AA and the lifetimes as long as 7 ps. We have also used the MSD method\cite{weitz} to estimate trap sizes,  and obtain results within a factor of $\sim 2$ from our simple geometrical approach.

\begin{table}
\caption{\label{tab:table4} Estimates for the trap size $r_{tr}$, trap lifetime $t_{tr}$, and the self-diffusion coefficient $D_{Ag}$ as a function of temperature T.}
\begin{ruledtabular}
\begin{tabular}{lccccccc}
  & T (K) & $r_{tr}$ (\AA) & $t_{tr}$ (ps) & $D_{Ag}$ ($cm^2/s$)\\  
\hline
$x=0.10$ & 640 & 0.81 & 5.89 &  \\
         & 800 & 1.41 & 3.31 & 1.29$\times$$10^{-7}$ \\
         & 1000 & 1.78 & 2.73 & 1.59$\times$$10^{-5}$ \\
\hline
\\
$x=0.15$ & 640 & 0.94 & 5.58 & 1.52$\times$$10^{-6}$ \\
         & 700 & 1.34 & 4.67 & 1.82$\times$$10^{-6}$ \\
         & 800 & 1.40 & 3.80 & 6.50$\times$$10^{-6}$ \\
         & 1000 & 1.84 & 2.66 & 2.06$\times$$10^{-5}$ \\
\end{tabular}
\end{ruledtabular}
\end{table} 

\begin{figure}
\includegraphics[width=3.0 in, height=3.0 in, angle=0]{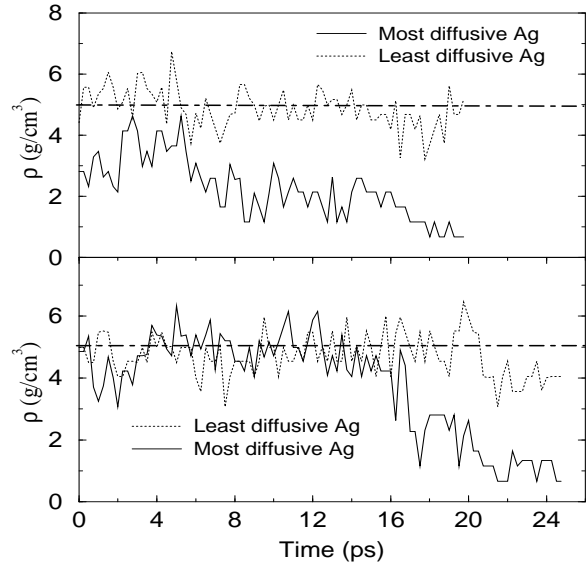}
\caption{Local density of the most and least mobile Ag atoms as a function of time in (GeSe$_3$)$_{0.90}$Ag$_{0.10}$ (top panel) and (GeSe$_3$)$_{0.85}$Ag$_{0.15}$ (bottom panel) glasses.\label{fig11}}
\end{figure}

\begin{figure}
\includegraphics[width=3.0 in, height=3.0 in, angle=0]{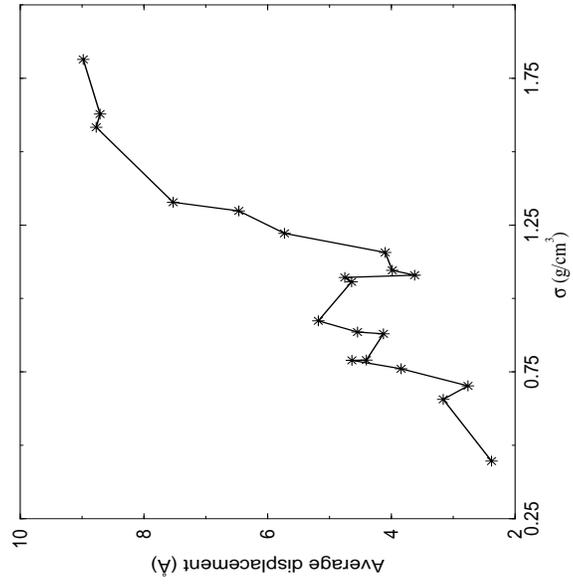}
\caption{Average displacement of the mobile Ag ions as a function of standard deviation of the local density in (GeSe$_3$)$_{0.90}$Ag$_{0.10}$ and (GeSe$_3$)$_{0.85}$Ag$_{0.15}$ glasses. The figure suggests that the more mobile Ag ions sample a wider range of local densities (see text).\label{fig12}}
\end{figure}

Further insight into the mechanism of diffusion in (Ge$_x$Se$_{1-x}$)$_{1-y}$Ag$_y$ can be found by studying the behavior of the molar volume of particular regions containing silver atoms. Hence we calculate the local density of the most and least mobile silver atoms as a function of time, then compare them to the density of the glass. To do so, we draw a sphere of radius R=4 \AA. The center of the sphere is the position of the Ag atoms we are tracking at a time t (the center of the sphere varies as a function of time). Then we calculate the mean density of atoms inside the sphere. Fig. \ref{fig11} illustrates the local density of a few Ag atoms as a function of time. As seen on the figure, the most mobile Ag atoms are consistently located in regions with a lower local density (lower local volume fraction) and higher disorder.  On the other hand, as we showed by direct calculation, there is little if any correlation between the trajectory-averaged mean density and the tendency to diffuse for the Ag atoms. On the other hand, perhaps unsurprisingly, we found also explored correlations between the average displacement of mobile Ag ions and the standard deviation of their local density ($\sigma_i$=$\sqrt{\langle\rho_{i}^{2}\rangle-\langle\rho_{i}\rangle^{2}}$, where $\rho_{i}$ is the local density of the Ag ion $i$, and $\langle \rangle$ means trajectory average.). We found that as the average displacement increases, the standard deviation becomes larger (see Fig. \ref{fig12}). The correlations are perhaps linear with much noise. This implies that the diffusive Ag ions are exploring a wide variety of densities and the weakly-diffusing Ag sample a restricted density range.

\section{Conclusion} 

We have presented \textit{ab initio} models of GeSe glasses heavily doped with Ag and studied the dynamics of the network with an emphasis on the motion of Ag ions. The models reproduce structural data, including reasonably subtle features in the diffraction data including the first peak (or shoulder) in S(Q). Wavelet methods help significantly in revealing intermediate range real-space correlations in the glasses. The atomistic motion of Ag$^+$ ions is detailed for short times with a reliable first principles interaction. We have shown by direct calculation that trapping centers exist, and have shown that local basis {\it ab initio} MD can provide direct insight into the processes of transition metal dynamics in amorphous chalcogenide materials. 

\section{Acknowledgments}

We thank the US National Science Foundation for support under grants DMR-0074624,  DMR-0205858 and DMR-0310933. We also gratefully acknowledge the support of Axon Technologies, Inc. We thank Dr. J. C. Phillips for many helpful comments and insights, and Drs. Jon Harrop,  S. N. Taraskin and Professor S. R. Elliott (University of Cambridge) for sharing their wavelet techniques for the analysis of intermediate range order.

\begin{thebibliography}{99}
\bibitem{gap} The optical gap is underestimated by nearly a factor of 1/2. This is an expected artifact of the Kohn-Sham eigenvalues.
\end{thebibliography}
\end{document}